# Dual electromagnetism: Helicity, spin, momentum, and angular momentum

Konstantin Y. Bliokh[1,2], Aleksandr Y. Bekshaev[3], and Franco Nori[1,4]

[1]*Advanced Science Institute, RIKEN, Wako-shi, Saitama 351-0198, Japan*
[2]*A. Usikov Institute of Radiophysics and Electronics, NASU, Kharkov 61085, Ukraine*
[3]*I. I. Mechnikov National University, Dvorianska 2, Odessa, 65082, Ukraine*
[4]*Physics Department, University of Michigan, Ann Arbor, Michigan 48109-1040, USA*

The *dual symmetry* between electric and magnetic fields is an important intrinsic property of Maxwell equations in free space. This symmetry underlies the conservation of optical *helicity*, and, as we show here, is closely related to the separation of *spin and orbital degrees of freedom* of light (the helicity flux coincides with the spin angular momentum). However, in the standard field-theory formulation of electromagnetism, the field Lagrangian is *not* dual symmetric. This leads to problematic dual-asymmetric forms of the *canonical* energy-momentum, spin, and orbital angular momentum tensors. Moreover, we show that the components of these tensors conflict with the helicity and energy conservation laws. To resolve this discrepancy between the symmetries of the Lagrangian and Maxwell equations, we put forward a dual-symmetric Lagrangian formulation of classical electromagnetism. This *dual electromagnetism* preserves the form of Maxwell equations, yields meaningful canonical energy-momentum and angular momentum tensors, and ensures a self-consistent separation of the spin and orbital degrees of freedom. This provides rigorous derivation of results suggested in other recent approaches. We make the Noether analysis of the dual symmetry and all the Poincaré symmetries, examine both local and integral conserved quantities, and show that only the dual electromagnetism naturally produces a complete self-consistent set of conservation laws. We also discuss the observability of physical quantities distinguishing the standard and dual theories, as well as relations to quantum weak measurements and various optical experiments.

## 1. Introduction

The symmetry between electric and magnetic fields in Maxwell's electromagnetism attracted considerable attention since the end of the 19$^{\text{th}}$ century [1,2]. It was then noticed that Maxwell equations in free space are symmetric with respect to the following exchange of the electric and magnetic fields:

$$\mathbf{E} \to \mathbf{B}, \quad \mathbf{B} \to -\mathbf{E}. \tag{1.1}$$

This discrete symmetry is a particular case of the continuous *dual symmetry* with respect to the electric-magnetic rotation

$$\begin{aligned} \mathbf{E} &\to \mathbf{E}\cos\theta + \mathbf{B}\sin\theta, \\ \mathbf{B} &\to \mathbf{B}\cos\theta - \mathbf{E}\sin\theta, \end{aligned} \tag{1.2}$$

where $\theta$ is an arbitrary pseudo-scalar. One of the manifestations of this symmetry is that all fundamental properties of free electromagnetic field (such as energy, momentum, angular momentum, etc.) are symmetric with respect to the transformation (1.2), which is referred to by Berry as "electric-magnetic democracy" [3].

Interest on the symmetry between the electric and magnetic properties of nature caught its second wind after seminal papers by Dirac that examined the possibility of the existence of magnetic charges (monopoles) [4]. Starting from the 1960s, this stimulated a series of works discussing dual-symmetric formulation of electromagnetism and light-matter interactions [5–21]. Simultaneously, in 1964 Lipkin discovered a series of novel conservation laws (sometimes called "Lipkin's zilches") in Maxwell equations, which were remarkably symmetric with respect to the electric and magnetic fields [22]. The Lipkin zilches include a pseudo-scalar, a pseudo-vector, and higher-rank tensors. Analysis of these conserved quantities revealed that the pseudo-scalar integrated over space is related to the difference between the numbers of right-hand and left-hand circularly-polarized photons, i.e., the optical *helicity* [23–28]. It was pointed out by Zwanziger [8] and also considered by Deser and Teitelboim [12] that it is the symmetry (1.2) that leads to the conservation of the helicity of light. However, in most other works the dual symmetry and helicity conservation were considered within different contexts.

Recently, the interest on helicity conservation raised again in connection with optical *chirality*, i.e., interaction of light with chiral structures [29–32]. Tang and Cohen [29,30] argued that the Lipkin's pseudo-scalar and pseudo-vector characterize the optical chirality density and its flux unrelated to the polarization of light. However, it was soon recovered in [33,34] that for monochromatic fields these quantities should be associated with the helicity density and its flux. Moreover, it was shown that the helicity flux represents the dual-symmetric *spin angular momentum* of light [33]. Thus, importantly, the dual symmetry and conservation of helicity are closely related to the definition of the spin density in the electromagnetic field – a longstanding problem by itself. Finally, this year, the fundamental relations between the dual symmetry, helicity conservation law, spin, and Lipkin's zilches were examined in detail in papers by Cameron, Barnett, and Yao [35]. Furthermore, Fernandez-Corbaton *et al.* [36] first considered the electric-magnetic symmetry and helicity conservation as a *practical* tool describing a number of experimentally-observed features in light-matter interactions.

Despite such extensive discussions about the dual symmetry (1.2) and helicity conservation, the standard electromagnetic field theory still has a significant drawback. Namely, the *Lagrangian of the electromagnetic field*, $\tilde{\mathcal{L}} = \left(E^2 - B^2\right)/2$, *is not dual-invariant* with respect to (1.2). This results in dual-*asymmetric* Noether currents and conservation laws [37,38]. In particular, the *canonical energy-momentum and angular-momentum tensors* are dual-asymmetric [37], which results in the known asymmetric definition of the *spin and orbital angular momenta* for the electromagnetic field [39]. (The usual symmetric energy-momentum tensor is obtained via an additional Belinfante symmetrization procedure, which is related to the separation of the spin and orbital degrees of freedom of the field [37,40].) Therefore, the helicity and spin densities become *inconsistent* with each other in the standard Lagrangian electromagnetic theory: the helicity flux does *not* coincide with the spin. Sometimes this evokes a false dual-asymmetric helicity (a partner of the asymmetric spin), which is *not conserved* in Maxwell equations [25,27,28,34] (see also [20], §1.6). Thus, there is a fundamental discrepancy in the symmetries of the free-space Lagrangian and field equations, which manifests itself in inconsistent definitions of the helicity, spin, and orbital angular momentum densities. Deser and Teitelboim [12] showed that even for the asymmetric Lagrangian, the integral action is dual-invariant, and the proper conserved helicity can be derived from it. However, this does not resolve the dual asymmetry of the spin and orbital quantities, and below we argue that the symmetry of the Lagrangian and the corresponding canonical Noether currents still makes an important difference. It is worth noticing that there were several attempts to restore the dual symmetry in the field Lagrangian [13–19] but all of these have difficulties and do not result in a clear, manifestly dual-symmetric formulation of electromagnetism.

In this paper we put forward a dual-symmetric modification of the free-space classical Maxwell electromagnetism. Our theory is based on a dual-symmetric Lagrangian which is invariant with respect to the transformation (1.2). Therefore, the straightforward application of the Noether theorem to this Lagrangian immediately yields the proper helicity conservation law.

The field equations, the symmetric energy-momentum tensor, and the corresponding angular-momentum tensor remain the same as in standard electromagnetic theory. At the same time, the *canonical* versions of these tensors (which directly follow from the Noether theorem but are usually considered as auxiliary quantities [37,38]) become meaningful physical quantities in our approach. They provide a dual-symmetric *separation of the spin and orbital degrees of freedom*, i.e., the spin and orbital parts of the momentum density (the Poynting vector) and the corresponding spin and orbital angular momenta of light. We show that only the dual-symmetric Lagrangian results in the canonical Noether currents – momentum, spin and orbital angular momenta, and boost momentum – *fully consistent* with each other and with the helicity conservation law. Thus, all discrepancies present in the standard electromagnetism are removed in our dual-symmetric theory. Note that the separation of spin and orbital parts of the angular momentum of light is a longstanding and controversial problem; it was recently re-examined by several authors [41–44]. In particular, the importance of the electric-magnetic symmetry was emphasized in [3,42,43]. Remarkably, the spin and orbital characteristics of the electromagnetic field, which were suggested in [3,42–44] using various arguments, now become *intrinsic* in the dual formulation of electromagnetism, and are derived in a rigorous way.

This paper is organized as follows. In Section 2 we summarize the standard Lagrangian formulation of classical electromagnetism, emphasizing its inherent difficulties related to its lack of duality. We present a complete Noether analysis of the dual symmetry and all Poincaré symmetries, examine both local and integral conserved quantities. Afterwards, in Section 3 we formulate the dual-symmetric modification of electromagnetic field theory, with its step-by-step comparison with Section 2 and other relevant results. To facilitate the comparison of the Noether currents in the standard and dual theories, we summarize them in Figure 1. Subsection 3.3 also contains a quantum-like operator representation of the conserved physical characteristics for monochromatic fields [3,42,44]; such self-consistent representation is impossible within standard electromagnetism. In Section 4 we discuss the issues of observability of spin and orbital angular momenta and local currents, which distinguish standard and dual theories. A number of known results and experiments are discussed from the viewpoint of dual symmetry. We argue that the suggested optical experiments are related to the concept of quantum weak measurements and unavoidably involve light-matter interactions. The remarks in Section 5 conclude the paper.

## 2. Standard electromagnetism

### *2.1. Field Lagrangian, Maxwell equations, and basic currents*

The electromagnetic field is described by the 4-potential $A^{\alpha}\left(r^{\alpha}\right)=\left(A^{0},\mathbf{A}\right)$ and antisymmetric field tensor

$$F^{\alpha\beta}=\partial^{\alpha}\wedge A^{\beta}=\left(\mathbf{E},\mathbf{B}\right). \tag{2.1}$$

Here and in what follows we use the Minkowski space-time $r^{\alpha}=\left(t,\mathbf{r}\right)$ with signature $\left(-,+,+,+\right)$ and assume natural units $\varepsilon_0=\mu_0=c=1$. We will adopt the transverse Coulomb gauge, $A^0=0$, $\nabla\cdot\mathbf{A}=0$ ($\mathbf{A}=\mathbf{A}_{\perp}$), because it is the transverse part of the potential which is gauge-invariant and can determine physically-meaningful quantities for the transverse radiation fields [39,41,43]. Also, all fields are assumed to be decaying sufficiently fast when $\left|\mathbf{r}\right|\to\infty$, to make all spatial integrals converging.

The standard Lagrangian of the free electromagnetic field is

$$\tilde{\mathcal{L}}=-\frac{1}{4}F^{\alpha\beta}F_{\alpha\beta}=\frac{1}{2}\left(E^2-B^2\right). \tag{2.2}$$

Here and in what follows, emphasizing the dual asymmetry of some important quantities, we will mark them with tilde; their dual-symmetric counterparts will be marked by the same letters without tilde. The field Hamiltonian, following from (2.2), is known to be [39] $\mathcal{H} = \frac{1}{2}\left(E^2 + B^2\right) = W$, where $W$ is the dual-symmetric energy density of the electromagnetic field.

The free-space Maxwell equations can be written as

$$\partial_\beta F^{\alpha\beta} = 0\,, \quad \partial_\beta {*F}^{\alpha\beta} = 0\,, \tag{2.3}$$

or

$$\nabla \cdot \mathbf{B} = \nabla \cdot \mathbf{E} = 0\,, \quad \partial_t \mathbf{E} = \nabla \times \mathbf{B}\,, \quad \partial_t \mathbf{B} = -\nabla \times \mathbf{E}\,. \tag{2.3'}$$

Here we introduced the *dual* field tensor:

$$*F^{\alpha\beta} \equiv \frac{1}{2}\varepsilon^{\alpha\beta\gamma\delta} F_{\gamma\delta} = \left(\mathbf{B}, -\mathbf{E}\right), \tag{2.4}$$

where $\varepsilon^{\alpha\beta\gamma\delta}$ is the Levi-Civita symbol. Note that only the first Maxwell equation in (2.3) represents the Euler-Lagrange equation of motion when varying the Lagrangian (2.2) with respect to $A^\alpha$. At the same time, the second equation in (2.3) is the Bianchi identity, which follows automatically from the form of the field tensor (2.1).

As compared to the field tensor (2.1), the dual field tensor (2.4) consists of the electric and magnetic fields $\mathbf{E}$ and $\mathbf{B}$ exchanged with each other via (1.1). Evidently, the Maxwell equations (2.3) are symmetric with respect to the dual exchange (1.1) $F^{\alpha\beta} \to *F^{\alpha\beta}$, because $**F^{\alpha\beta} = -F^{\alpha\beta}$. It is easy to show that equations (2.3') are also invariant with respect to the continuous dual transformations (1.2).

Note that the coupling with matter (which consists of *electric* charges and currents) *breaks* the dual symmetry of the free field. The Lagrangian (2.2) acquires the coupling part $\tilde{\mathcal{L}}_C = A^\alpha j_{E\alpha}$, where $j_E^\alpha$ is the electric 4-current, and the first equation of motion (2.3) becomes $\partial_\beta F^{\alpha\beta} = j_E^\alpha$. It is seen from this that the presence of both electric and magnetic charges is problematic in this picture, because it requires a modification of the second equation (2.3) with the magnetic current $j_M^\alpha$: $\partial_\beta *F^{\alpha\beta} = j_M^\alpha$ [4–20]. However, this would contradict the Bianchi identity for the field tensor (2.1), and the relation between the field and potential should be modified.

In contrast to Maxwell equations, the field Lagrangian (2.2) is *not* invariant with respect to the duality transformations (1.1) and (1.2). To illustrate its transformation properties, we introduce the complex *Riemann-Silberstein vector* $\mathbf{D} = \mathbf{E} + i\mathbf{B}$ [45]. The Lorentz transformations represent complex-angle rotations of this vector [38], and its square, $\mathbf{D}\cdot\mathbf{D} = \left(E^2 - B^2\right) + 2i\,\mathbf{E}\cdot\mathbf{B} \equiv I_1 + i\,I_2$, provides the two Lorentz invariants of the electromagnetic field:

$$\begin{aligned} I_1 &= -\frac{1}{2}F^{\alpha\beta}F_{\alpha\beta} = \left(E^2 - B^2\right), \\ I_2 &= -\frac{1}{2}*F^{\alpha\beta}F_{\alpha\beta} = 2\mathbf{E}\cdot\mathbf{B}\,. \end{aligned} \tag{2.5}$$

In terms of the Riemann-Silberstein vector, the dual transformation (1.2) becomes the $U(1)$ gauge transformation $\mathbf{D} \to e^{-i\theta}\mathbf{D}$, whereas the field Lagrangian (2.2) takes the form $\tilde{\mathcal{L}} = \frac{1}{2}\mathrm{Re}\left(\mathbf{D}\cdot\mathbf{D}\right) = \frac{1}{2}I_1$. From these equations we immediately see that the dual transformation (1.2) induces a rotation of the field invariants

$$\begin{aligned} I_1 &\to I_1 \cos 2\theta + I_2 \sin 2\theta\,, \\ I_2 &\to I_2 \cos 2\theta - I_1 \sin 2\theta\,, \end{aligned} \tag{2.6}$$

and the Lagrangian is transformed as

$$\tilde{\mathcal{L}} \to \frac{1}{2}\left(I_1 \cos 2\theta + I_2 \sin 2\theta\right) = \tilde{\mathcal{L}} \cos 2\theta - \frac{1}{4} {*F^{\alpha\beta}} F_{\alpha\beta} \sin 2\theta . \tag{2.7}$$

The transformation (2.7) changes the Lagrangian, but does not affect the Maxwell equations of motion. Indeed, ${*F^{\alpha\beta}} F_{\alpha\beta} = 2\partial_\alpha \left({*F^{\alpha\beta}} A_\beta\right)$ [we can use here $\partial_\alpha {*F^{\alpha\beta}} = 0$, because this is the Bianchi identity rather than the equation of motion], and the transformation (2.7) adds a total divergence alongside with a multiplication by the constant $\cos 2\theta$.

Considering an infinitesimal version of (2.7) with $\theta \to 0$, we have

$$\tilde{\mathcal{L}} \to \tilde{\mathcal{L}} - \theta \partial_\alpha \left({*F^{\alpha\beta}} A_\beta\right). \tag{2.8}$$

From here it is speculated in [20], §1.6, that the dual symmetry of Maxwell equations implies the conservation of the *dual current*

$$\tilde{J}^\alpha = {*F^{\alpha\beta}} A_\beta, \quad \text{i.e.,} \quad \tilde{J}^0 \equiv \tilde{H} = \mathbf{A} \cdot \mathbf{B}, \quad \tilde{\mathbf{J}} \equiv \tilde{\mathbf{S}} = \mathbf{E} \times \mathbf{A} . \tag{2.9}$$

The components of the current (2.9) are often considered as the *helicity density* $\tilde{H}$ and the *spin angular momentum* density $\tilde{\mathbf{S}}$ (coinciding here with the helicity flux density) of the field [25,34,39,41]. However, this is not the case, and this false helicity is *not conserved* [25–28]:

$$\partial_\alpha \tilde{J}^\alpha = -I_2 \neq 0, \quad \int \tilde{H}\, dV \neq \text{const}, \tag{2.10}$$

where $dV \equiv d^3\mathbf{r}$ and "const" denotes constancy in time. It was shown in a number of works [27,28,35,43] that the dual-asymmetric definitions for the helicity and spin densities (2.9) are not satisfactory in the general case. Instead, the dual-symmetric modifications of Eqs. (2.9) and (2.10) form the true *helicity-spin current* providing the helicity conservation [8,12,16,27,28,35]:

$$J^0 \equiv H = \frac{1}{2}\left(\mathbf{A} \cdot \mathbf{B} - \mathbf{C} \cdot \mathbf{E}\right), \quad \mathbf{J} \equiv \mathbf{S} = \frac{1}{2}\left(\mathbf{E} \times \mathbf{A} + \mathbf{B} \times \mathbf{C}\right). \tag{2.11}$$

$$\partial_\alpha J^\alpha = 0, \quad \int H\, dV = \text{const}. \tag{2.12}$$

Here $\mathbf{A}$ and $\mathbf{C}$ are the magnetic and electric vector-potentials: $\nabla \times \mathbf{C} = -\mathbf{E}$ and $\nabla \times \mathbf{A} = \mathbf{B}$ [see Eqs. (3.6) below]. The second quantity in (2.11) was recently identified as the physically-meaningful spin angular momentum density [42–44].

Let us make one more observation, the meaning of which will be clarified below. Akin to the infinitesimal dual transformation, we consider an infinitesimal time-translation transformation: $t \to t + \tau$, $\tau \to 0$. In a way similar to Eq. (2.8), using Maxwell equation $\partial_\alpha F^{\alpha\beta} = 0$, one can show that the value of the Lagrangian (2.2) on the minimal-action trajectories is transformed as

$$\tilde{\mathcal{L}} \to \tilde{\mathcal{L}} - \tau \partial_\alpha \left(F^{\alpha\beta} \partial_t A_\beta\right). \tag{2.13}$$

Then, akin to Eqs. (2.9) and (2.10), this evokes a false *energy-momentum current*:

$$\tilde{P}^\alpha = -F^{\alpha\beta} \partial_t A_\beta, \quad \text{i.e.,} \quad \tilde{P}^0 \equiv \tilde{W} = E^2, \quad \tilde{\mathbf{P}} = \mathbf{P} = \mathbf{E} \times \mathbf{B}, \tag{2.14}$$

where the energy is *not conserved*:

$$\partial_\alpha \tilde{P}^\alpha = \frac{1}{2} \partial_t I_1 \neq 0, \quad \int \tilde{W}\, dV \neq \text{const}. \tag{2.15}$$

The current (2.14) and (2.15) contains the proper Poynting vector ($=$ energy flux density $=$ momentum density) $\mathbf{P} = \mathbf{E} \times \mathbf{B}$, but it also contains the dual-asymmetric energy density $\tilde{W} = E^2$, which would appear if the energy would be concentrated solely in the electric field. Of course, the true conserved energy-momentum current, associated with the time-translation invariance, is dual-symmetric and reads

$$P^0 \equiv W = \frac{1}{2}\left(E^2 + B^2\right), \quad \mathbf{P} = \mathbf{E} \times \mathbf{B}, \tag{2.16}$$

$$\partial_\alpha P^\alpha = 0, \quad \int W\, dV = \text{const}. \tag{2.17}$$

The proper energy conservation (2.16) and (2.17) follows from the Noether-theorem analysis, which is considered in Section 2.2 below. Also, Deser and Teitelboim [12] showed that the proper helicity conservation (2.11) and (2.12) can be obtained from the dual-asymmetric Lagrangian (2.2). However, we will also see that the *false* spin density (2.9) $\tilde{\mathbf{S}} = \mathbf{E} \times \mathbf{A}$ and the *false* energy density (2.14) $\tilde{W} = E^2$ appear in the components of the *canonical angular momentum tensor* following from the Lagrangian (2.2). This is an important discrepancy of the standard dual-asymmetric Lagrangian formulation of electromagnetism which fails to produce conservation laws consistent with each other.

### *2.2. Energy-momentum and angular momentum tensors.*

In this Section we summarize the main conservation laws for the free electromagnetic field, which are associated with the Poincaré group of symmetries, i.e., translations and rotations of the Minkowski space-time. We will obtain the corresponding conserved quantities by applying the Noether theorem to the Lagrangian (2.2), and will emphasize their specific features which are important for our theory. This Section mostly follows the standard textbook approach [37,38].

*2.2.1. Canonical tensors.* The invariance with respect to translations in space and time generates the conservation of momentum and energy. The application of the Noether theorem [46] yields the following dual-asymmetric but conserved energy-momentum tensor from the Lagrangian (2.2):

$$\tilde{T}^{\alpha\beta} = \left(\partial^\alpha A^\gamma\right) F^\beta{}_\gamma - \frac{1}{4} g^{\alpha\beta} F^{\gamma\delta} F_{\gamma\delta}, \quad \partial_\beta \tilde{T}^{\alpha\beta} = 0. \qquad (2.18)$$

where $g^{\alpha\beta}$ is the metric tensor. Tensor (2.18) is *non-symmetric*, $\tilde{T}^{\alpha\beta} \neq \tilde{T}^{\beta\alpha}$, and is known as *canonical* energy-momentum tensor. The corresponding 4-momentum density of the field is given by $\tilde{P}_O^\alpha = \tilde{T}^{\alpha 0}$ (the indices "*O*" and "*S*" indicate "orbital" and "spin" quantities, see below). Its temporal component (in the chosen Coulomb gauge) reads

$$\tilde{P}_O^0 = -\mathbf{E} \cdot \partial_t \mathbf{A} + \frac{1}{2}\left(B^2 - E^2\right) = \frac{1}{2}\left(E^2 + B^2\right) = W, \qquad (2.19)$$

which is the proper dual-symmetric energy density (2.16). At the same time, the spatial components of the canonical momentum density, form the dual-asymmetric vector

$$\tilde{\mathbf{P}}_O = \mathbf{E} \cdot (\nabla) \mathbf{A}. \qquad (2.20)$$

(Here and in what follows we use the notations by Berry [3], for which the scalar product links the vectors $\mathbf{E}$ and $\mathbf{A}$, whereas the gradient is external: $\tilde{P}_{Oi} = E_j \nabla_i A_j$; the Latin indices $i, j, k$ take on values 1,2,3.) As we will see, the canonical momentum density $\tilde{\mathbf{P}}_O$ represents the *orbital* part of the energy flux density. The total energy flux density (Poynting vector) $\mathbf{P}$ is given by the spatial components of $\tilde{T}^{0\alpha} = P^\alpha$, which form the proper conserved energy-momentum current (2.16) in the chosen Coulomb gauge.

Next, the symmetry with respect to Lorentz transformations (rotations of the Minkowski space-time) results in the conservation of the relativistic angular momentum. Recall that the angular momentum is described in relativistic field theory by a rank-3 tensor $M^{\alpha\beta\gamma}$, where the anti-symmetric rank-2 tensor $M^{\alpha\beta 0}$ can be represented by a pair of 3-vectors: $M^{\alpha\beta 0} = (\mathbf{N}, \mathbf{M})$. Here the pseudo-vector $M_i = \frac{1}{2}\varepsilon_{ijk} M^{jk0}$ is the usual angular momentum, which has the form of $\mathbf{M} = \mathbf{r} \times \mathbf{P}$ for point particles and is related to the symmetry with respect to spatial rotations. At the same time, the vector $N_i = M^{0i0}$ could be referred to as the *boost momentum*. It is related to the symmetry with respect to Lorentz boosts and has the form $\mathbf{N} = t\,\mathbf{P} - \mathbf{r}\,W$ for point particles. Conservation of the boost momentum ensures *the rectilinear motion of the energy centroid* in free space [38,47]. The application of the Noether theorem to the Lagrangian (2.2) results in the following dual-asymmetric but conserved *canonical* angular-momentum tensor:

$$\tilde{M}^{\alpha\beta\gamma} = r^\alpha \tilde{T}^{\beta\gamma} - r^\beta \tilde{T}^{\alpha\gamma} + \tilde{S}^{\alpha\beta\gamma} \equiv \tilde{L}^{\alpha\beta\gamma} + \tilde{S}^{\alpha\beta\gamma}, \quad \partial_\gamma \tilde{M}^{\alpha\beta\gamma} = 0. \tag{2.21}$$

Here $\tilde{S}^{\alpha\beta\gamma}$ is the *spin tensor*, which has the form

$$\tilde{S}^{\alpha\beta\gamma} = F^{\gamma\alpha} A^\beta - F^{\gamma\beta} A^\alpha, \quad \partial_\gamma \tilde{S}^{\alpha\beta\gamma} = \tilde{T}^{\alpha\beta} - \tilde{T}^{\beta\alpha} \neq 0. \tag{2.22}$$

Importantly, the canonical angular momentum tensor (2.21) suggests a natural *separation into orbital and spin parts*, $\tilde{L}^{\alpha\beta\gamma}$ and $\tilde{S}^{\alpha\beta\gamma}$. Calculating the angular-momentum vector $\tilde{\mathbf{M}}$ from Eqs. (2.18)–(2.22), we arrive at

$$\tilde{\mathbf{M}} = \mathbf{E} \cdot (\mathbf{r} \times \nabla)\mathbf{A} + \mathbf{E} \times \mathbf{A} \equiv \tilde{\mathbf{L}} + \tilde{\mathbf{S}}. \tag{2.23}$$

Equation (2.23) describes the canonical angular momentum density, where the orbital angular momentum can be written in 'mechanical' form $\tilde{\mathbf{L}} = \mathbf{r} \times \tilde{\mathbf{P}}_O$, consistent with the canonical momentum density (2.20). At the same time, the expression for the spin angular momentum density $\tilde{\mathbf{S}} = \mathbf{E} \times \mathbf{A}$ here coincides with the *false* (non-conserved) dual-asymmetric helicity flux density (2.9) rather than with the proper (conserved) dual-symmetric quantity (2.11) $\mathbf{S}$! This is the first important discrepancy in the standard Lagrangian formulation of electromagnetism.

Next, the canonical boost momentum density $\tilde{\mathbf{N}}$ derived from tensor (2.21) can also be separated into orbital and spin parts, which yields

$$\tilde{\mathbf{N}} = \mathbf{E} \cdot (t\nabla + \mathbf{r}\partial_t)\mathbf{A} - A^0 \mathbf{E} \equiv \tilde{\mathbf{N}}_O + \tilde{\mathbf{N}}_S. \tag{2.24}$$

In the chosen Coulomb gauge, this results in

$$\tilde{\mathbf{N}}_O = t\,\tilde{\mathbf{P}}_O - \mathbf{r}\tilde{W}, \quad \tilde{\mathbf{N}}_S = 0. \tag{2.25}$$

Thus, the spin part of the boost momentum vanishes [43] (apparently, this reflects the truly *intrinsic* nature of spin, which does not involve energy transport [42,48–53]). At the same time, the orbital boost momentum $\tilde{\mathbf{N}}_O = \tilde{\mathbf{N}}$ takes a clear mechanical-like form in (2.25). It involves the canonical momentum density (2.20) $\tilde{\mathbf{P}}_O$ and *false* (non-conserved) dual-asymmetric energy density (2.14) $\tilde{W} = -\mathbf{E} \cdot \partial_t \mathbf{A} = E^2$ rather than the proper conserved quantity (2.16) and (2.19) $W$! And this is the second important discrepancy in the standard Lagrangian formulation of electromagnetism.

*2.2.2. Symmetrized tensors.* Although the canonical energy-momentum tensor (2.18) and momentum density (2.20), directly follow from the Noether theorem and Lagrangian (2.2), they are usually considered as auxiliary quantities. Instead the *symmetric* energy-momentum tensor $\mathcal{T}^{\alpha\beta} = \mathcal{T}^{\beta\alpha}$ is obtained via the Belinfante symmetrization procedure [40], i.e., the addition of the suitable total divergence to the canonical energy-momentum tensor. As applied to the tensor $\tilde{T}^{\alpha\beta}$, this procedure results in:

$$\mathcal{T}^{\alpha\beta} = \tilde{T}^{\alpha\beta} + \partial_\gamma \tilde{K}^{\alpha\beta\gamma} = F^{\alpha\gamma} F^\beta{}_\gamma - \frac{1}{4} g^{\alpha\beta} F^{\gamma\delta} F_{\gamma\delta}, \quad \partial_\beta \mathcal{T}^{\alpha\beta} = 0, \tag{2.26}$$

where the tensor $\tilde{K}^{\alpha\beta\gamma}$ is constructed from the spin tensor (2.22):

$$\tilde{K}^{\alpha\beta\gamma} = \frac{1}{2}\left(\tilde{S}^{\beta\gamma\alpha} + \tilde{S}^{\alpha\gamma\beta} - \tilde{S}^{\alpha\beta\gamma}\right) = -A^\alpha F^{\beta\gamma}. \tag{2.27}$$

When integrated over the whole space, both tensors (2.18) and (2.26) yield the same momentum of the field (see Section 2.2.3 below). The symmetric tensor (2.26) is manifestly gauge-invariant, and its components contain: The proper energy density $W$, the momentum density coincident with the total energy flux (Poynting vector) $\mathbf{P}$, and also the Maxwell stress tensor $\sigma_{ij}$ for the electromagnetic field:

$$\mathcal{T}^{00} = W = \frac{1}{2}\left(E^2 + B^2\right), \quad \mathcal{T}^{i0} = P_i = (\mathbf{E} \times \mathbf{B})_i, \tag{2.28}$$

$$\mathcal{T}^{ij} = -\sigma_{ij} = -\left(E_i E_j + B_i B_j - \delta_{ij} W\right). \tag{2.29}$$

Importantly, the quantities (2.26), (2.28), and (2.29) are dual-invariant with respect to the transformation (1.2), *in contrast* to the canonical quantities (2.18) and (2.20).

The symmetrization procedure (2.26) has an important physical meaning [37,48–51]. Let us consider the tensor $\partial_\gamma \tilde{K}^{\alpha\beta\gamma} = -F^{\beta\gamma}\partial_\gamma A^\alpha$, which is added to the canonical energy-momentum tensor. Its contribution to the field momentum is $\tilde{P}_{Si} = \partial_\gamma \tilde{K}^{i0\gamma}$:

$$\tilde{\mathbf{P}}_S = -(\mathbf{E}\cdot\nabla)\mathbf{A}. \tag{2.30}$$

This is the *spin current*, i.e., the spin part of the energy flux density. Indeed, the Poynting vector (2.28) is the sum of the orbital part (2.20) and spin part (2.30):

$$\mathbf{P} = \tilde{\mathbf{P}}_O + \tilde{\mathbf{P}}_S. \tag{2.31}$$

Although the spin current makes no contribution to the integral momentum of the field,

$$\int \tilde{\mathbf{P}}_S \, dV = 0, \tag{2.32}$$

circulations of the orbital and spin energy fluxes produce, respectively, orbital and spin angular momenta (2.23):

$$\tilde{\mathbf{L}} = \mathbf{r}\times\tilde{\mathbf{P}}_O, \quad \int \tilde{\mathbf{S}}\, dV = \int \mathbf{r}\times\tilde{\mathbf{P}}_S \, dV. \tag{2.33}$$

Here the second equation is obtained using integration by parts with $\nabla\cdot\mathbf{E} = 0$, and it displays a "nonlocal" action of the spin current. (This peculiar current does not transport energy and flows along the boundary resembling the magnetization current or topological quantum-Hall states in condensed-matter systems.) A detailed description and analysis of the spin and orbital energy fluxes can be found in [3,37,42,48–53]. Note also that an analogue of the second equation (2.33) for the spin part of the boost momentum $\tilde{\mathbf{N}}_S$ reads $\int \tilde{\mathbf{N}}_S dV = \int t\,\tilde{\mathbf{P}}_S dV = 0$, and does not give new information.

A *symmetrized* angular momentum tensor can be constructed from the symmetric energy-momentum tensor (2.26). Since the latter includes *both* orbital and spin parts of the momentum density, the angular momentum tensor becomes

$$\mathcal{M}^{\alpha\beta\gamma} = r^\alpha \mathcal{T}^{\beta\gamma} - r^\beta \mathcal{T}^{\alpha\gamma}, \quad \partial_\gamma \mathcal{M}^{\alpha\beta\gamma} = 0. \tag{2.34}$$

This tensor is dual symmetric, and it differs from the canonical angular momentum tensor (2.21): $\mathcal{M}^{\alpha\beta\gamma} \neq \tilde{M}^{\alpha\beta\gamma}$. Nonetheless, when integrated over space, both tensors $\tilde{M}^{\alpha\beta\gamma}$ and $\mathcal{M}^{\alpha\beta\gamma}$ yield the same total angular momentum of the field [37] (see Section 2.2.3). The angular momentum and boost momentum densities following from the tensor (2.34) are expressed through the symmetrized energy-momentum components (2.28) [cf. Eqs. (2.23)−(2.25)]:

$$\boldsymbol{\mathcal{M}} = \mathbf{r}\times\mathbf{P}, \quad \boldsymbol{\mathcal{N}} = t\,\mathbf{P} - \mathbf{r}W. \tag{2.35}$$

We emphasize that $\boldsymbol{\mathcal{M}}$ describes the *total* angular momentum density of the field, without separation of the orbital and spin parts.

Thus, the symmetrized energy-momentum and angular-momentum tensors (2.26) and (2.34) are convenient characteristics for calculations of the integral (non-local) dynamical properties of the field. However, it is *impossible* to separate spin and orbital angular-momentum degrees of freedom in $\mathcal{T}^{\alpha\beta}$ and $\mathcal{M}^{\alpha\beta\gamma}$ without involving the canonical tensors.

*2.2.3. Integral conserved quantities.* Above we considered *local* conservation laws associated with the dual and Poincaré symmetries. It is also interesting to discuss the corresponding conserved *integral* (i.e., non-local) quantities.

First, Eqs. (2.9)−(2.12) and (2.14)−(2.17) involving true and false helicity, spin, and energy densities can be formulated in the following integral form:

$$\text{const} \neq \int \tilde{H}\, dV \neq \int H\, dV = \text{const}, \tag{2.36}$$

$$\text{const} \neq \int \tilde{W} dV \neq \int W dV = \text{const}. \tag{2.37}$$

Next, from Eqs. (2.18) and (2.26), the integral energy-momentum following from the canonical and symmetric energy-momentum tensors are both conserved in time and equal to each other:

$$\int \tilde{T}^{\alpha 0} dV = \int \mathcal{T}^{\alpha 0} dV = \text{const}\,, \tag{2.38}$$

i.e.,

$$\int W\, dV = \text{const}\,, \quad \int \tilde{\mathbf{P}}_O\, dV = \int \mathbf{P}\, dV = \text{const}\,. \tag{2.39}$$

The second equality here is consistent with the vanishing of the integral spin momentum, Eqs. (2.31) and (2.32).

Similar equalities take place for the angular momenta following from canonical and symmetrized angular-momentum tensors (2.21) and (2.34):

$$\int \tilde{M}^{\alpha\beta 0} dV = \int \mathcal{M}^{\alpha\beta 0} dV = \text{const}\,. \tag{2.40}$$

In components, this yields

$$\int \tilde{\mathbf{M}}\, dV = \int \boldsymbol{\mathcal{M}}\, dV = \text{const}\,, \quad \int \tilde{\mathbf{N}}\, dV = \int \tilde{\mathbf{N}}_O dV = \int \boldsymbol{\mathcal{N}}\, dV = \text{const}\,, \tag{2.40'}$$

i.e.,

$$\int \left(\mathbf{r}\times\tilde{\mathbf{P}}_O + \tilde{\mathbf{S}}\right) dV = \int \left(\mathbf{r}\times\mathbf{P}\right) dV = \text{const}\,, \tag{2.41}$$

$$\int \left(t\,\tilde{\mathbf{P}}_O - \mathbf{r}\tilde{W}\right) dV = \int \left(t\,\mathbf{P} - \mathbf{r}W\right) dV = \text{const}\,. \tag{2.42}$$

Interesting new equality follows from Eqs. (2.42) and (2.39):

$$\int \mathbf{r}\tilde{W} dV = \int \mathbf{r} W dV\,, \;\; \text{i.e.,} \;\; \int \mathbf{r} E^2 dV = \int \mathbf{r} B^2 dV\,, \tag{2.43}$$

although $\int \tilde{W} dV \neq \int W dV$ and $\int E^2 dV \neq \int B^2 dV$.

Equation (2.42) results in the equation of the rectilinear motion of the centroid of the energy density $W$. Indeed, using $\partial_t\left(\int \mathbf{r} W dV - \int t\,\mathbf{P}\, dV\right) = 0$ and the energy-momentum conservation (2.39), we obtain

$$\partial_t \mathbf{R}_W = \frac{\int \mathbf{P}\, dV}{\int W dV} = \text{const}\,, \quad \mathbf{R}_W \equiv \frac{\int \mathbf{r} W dV}{\int W dV}\,. \tag{2.44}$$

At the same time, we note that the centroid of the false energy density $\tilde{W} = E^2$ entering the canonical boost-momentum density $\tilde{\mathbf{N}}$ does not move rectilinearly, because $\tilde{W}$ is not conserved, Eq. (2.37):

$$\text{const} \neq \partial_t \tilde{\mathbf{R}}_W \neq \frac{\int \tilde{\mathbf{P}}_O\, dV}{\int \tilde{W}\, dV} \neq \text{const}\,, \quad \tilde{\mathbf{R}}_W \equiv \frac{\int \mathbf{r}\tilde{W}\, dV}{\int \tilde{W}\, dV}\,. \tag{2.45}$$

Thus, the equation of motion of the energy centroid does not follow from the form of the canonical boost momentum (2.24) and (2.25) $\tilde{\mathbf{N}}$ in standard electromagnetism.

Equations (2.36)–(2.44) summarize all non-local conserved quantities following from the differential conservation laws. In addition, there are also important integral conserved quantities which have *no* differential counterparts. Namely, despite the fact that the spin and orbital parts of the angular momentum density (2.21) do not form conserved Noether currents, see Eq. (2.22), their integral values are conserved in time [41]:

$$\int \tilde{L}^{\alpha\beta 0} dV = \text{const}\,, \quad \int \tilde{S}^{\alpha\beta 0} dV = \text{const}\,, \tag{2.46}$$

i.e.,

$$\int \tilde{\mathbf{L}}\, dV = \text{const}\,, \quad \int \tilde{\mathbf{S}}\, dV = \text{const}\,. \tag{2.46'}$$

This can be verified by substituting here the spin and orbital angular momentum densities (2.23) and performing differentiation with respect to time. Equations (2.46') agree with the approach of modern optics, where the spin and orbital angular momenta of light are regarded as *separately observable* and *conserved* (in free space) quantities [41,54] (for reviews, see [55]). Note, however, that the integral values of the spin and orbital angular momenta $\tilde{\mathbf{L}}$ and $\tilde{\mathbf{S}}$ are *not* dual-symmetric. Moreover, their dual-symmetric versions are also conserved quantities. For instance, the dual-symmetric spin $\mathbf{S}$, entering the proper helicity conservation (2.11) and (2.12), also obeys

$$\int \mathbf{S}\, dV = \text{const}\,, \quad \text{but} \quad \int \mathbf{S}\, dV \neq \int \tilde{\mathbf{S}}\, dV\,. \tag{2.47}$$

Equations (2.36), (2.37), (2.45), and (2.47) demonstrate that the spin and orbital angular momenta following from the canonical angular-momentum tensor (2.21) and standard Lagrangian (2.2) conflict with the helicity and energy conservations not only locally but also in their integral values.

*2.2.4. Summary of the problems.* The above picture shows that the canonical energy-momentum and angular-momentum tensors are intimately related to the *separation of the spin and orbital degrees of freedom* of the electromagnetic field. In turn, the properly defined spin is a crucial part of the helicity conservation (2.11) and (2.12). Therefore, it is important that these canonical tensors $\tilde{T}^{\alpha\beta}$ and $\tilde{M}^{\alpha\beta\gamma}$ should be physically meaningful and *consistent with each other and the helicity conservation*. However, in the standard formulation, they lack the dual invariance with respect to the transformation (1.2). As a result of this, the Lagrangian (2.2) leads to the false spin density (2.9) $\tilde{\mathbf{S}}$ and false energy density (2.14) $\tilde{W}$, which appear in the components of $\tilde{M}^{\alpha\beta\gamma}$; while the proper dual-symmetric spin and energy densities, $\mathbf{S}$ and $W$, are given by Eqs. (2.11) and (2.16) or (2.28). In turn, the definition of the spin is closely related to the definitions of other quantities: orbital angular momentum density $\tilde{\mathbf{L}}$, spin and orbital parts of the energy flux (momentum) density, $\tilde{\mathbf{P}}_O$ and $\tilde{\mathbf{P}}_S$. Recent investigations suggested that all these dual-asymmetric quantities should be substituted by their dual-symmetric modifications [3,42–44,51,53]. Thus, we observe a number of inconsistencies in the canonical conservation laws of standard electromagnetic theory; these are all summarized in Figure 1. We have also shown that the discrepancies between the dual-symmetric and asymmetric versions of the helicity, energy, spin and orbital angular momenta appear in both local and integral values of these quantities.

It should be emphasized that the above discrepancies involve *measurable* quantities. There are no doubts that the integral helicity and energy are observable. Modern optics regards the spin and orbital parts of the angular momentum as separately measurable quantities [41,54,55]. Remarkably, even *local* densities of energy, spin, and orbital energy flow can be retrieved from optical experiments [51,54,56]. We further discuss the observability issues in Section 4.

## 3. Dual electromagnetism

### *3.1.Dual-symmetric Lagrangian and basic currents.*

To construct the Lagrangian formalism which would contain the dual symmetry (1.2), we first consider the field tensor $F^{\alpha\beta}$ and its dual pseudo-tensor $*F^{\alpha\beta} = G^{\alpha\beta}$ as *independent* quantities based on two different 4-potentials $A^\alpha$ and $C^\alpha$:

$$F^{\alpha\beta} = \partial^\alpha \wedge A^\beta = (\mathbf{E}, \mathbf{B})\,, \quad G^{\alpha\beta} = \partial^\alpha \wedge C^\beta = (\mathbf{B}, -\mathbf{E})\,, \tag{3.1}$$

Such two-potential representation is extensively used in works about magnetic monopoles and optical helicity [5–21,23–28,35]. Next, we suggest a *dual-symmetric Lagrangian*, which is the symmetrization of the standard Lagrangian (2.2) with respect to $F^{\alpha\beta}$ and $G^{\alpha\beta}$:

$$\mathcal{L} = -\frac{1}{8}\left(F^{\alpha\beta}F_{\alpha\beta} + G^{\alpha\beta}G_{\alpha\beta}\right). \tag{3.2}$$

Here the two fields are considered independently, but since they actually describe the *same* electromagnetic field, we should impose an additional *duality constraint*:

$$*F^{\alpha\beta} = G^{\alpha\beta}, \quad \text{or} \quad *\left(\partial^\alpha \wedge A^\beta\right) = \left(\partial^\alpha \wedge C^\beta\right). \tag{3.3}$$

The value of the Lagrangian vanishes with this constraint: $\mathcal{L} = 0$. Nonetheless, by varying the Lagrangian (3.2) with respect to the potentials $A^\alpha$ and $C^\alpha$ independently, we obtain two Euler-Lagrange equations of motion:

$$\partial_\beta F^{\alpha\beta} = 0, \quad \partial_\beta G^{\alpha\beta} = 0. \tag{3.4}$$

At the same time, the Bianchi identity for the fields (3.1) yields

$$\partial_\beta *F^{\alpha\beta} = 0, \quad \partial_\beta *G^{\alpha\beta} = 0. \tag{3.5}$$

Applying now the constraint (3.3), we see that the two pairs of equations (3.4) and (3.5) result in the same pair of free-space Maxwell equations (2.3). This derivation of the equations of motion is equivalent to the use of a Lagrange multiplier with the constraint (3.3) in the Lagrangian.

Thus, the Maxwell equations are recovered from the symmetrized Lagrangian (3.2). It should be noticed, however, that this formalism describes only *free* radiation transverse fields. As it was pointed out in Section 2.1, a coupling with matter brings about currents in the right-hand side of Eqs. (3.4), which would contradict Eqs. (3.5). In this case, one has to modify the relation (3.1) between the potentials and fields, to include the longitudinal parts of the fields [4–20].

As in Section 2, we now consider only the transverse radiation fields, and adopt the transverse Coulomb gauge for the potentials: $A^0 = C^0 = 0$ and $\nabla \cdot \mathbf{A} = \nabla \cdot \mathbf{C} = 0$ ($\mathbf{A} = \mathbf{A}_\perp$, $\mathbf{C} = \mathbf{C}_\perp$) [35,43]. Note that the Coulomb gauge and constraint (3.3) are equivalent to "Maxwell equations" for the vector-potentials [35]:

$$\nabla \cdot \mathbf{A} = \nabla \cdot \mathbf{C} = 0, \quad \partial_t \mathbf{A} = \nabla \times \mathbf{C}\left(= -\mathbf{E}\right), \quad \partial_t \mathbf{C} = -\nabla \times \mathbf{A}\left(= -\mathbf{B}\right). \tag{3.6}$$

Therefore, the duality rotation (1.2) generates the same rotation of the vector-potentials [8,10,13,17,18,35]:

$$\begin{aligned} \mathbf{A} &\to \mathbf{A}\cos\theta + \mathbf{C}\sin\theta, \\ \mathbf{C} &\to \mathbf{C}\cos\theta - \mathbf{A}\sin\theta. \end{aligned} \tag{3.7}$$

The dual-symmetric formalism acquires a particularly laconic form if we introduce the *complex* Riemann-Silberstein-like 4-potential $\mathrm{X}^\alpha$ and the corresponding field tensor $D^{\alpha\beta}$:

$$\mathrm{X}^\alpha = A^\alpha + i\,C^\alpha, \quad D^{\alpha\beta} = F^{\alpha\beta} + i\,G^{\alpha\beta} = \partial^\alpha \wedge \mathrm{X}^\beta = \left(\mathbf{D}, -i\mathbf{D}\right). \tag{3.8}$$

In this manner, the Lagrangian (3.2) becomes

$$\mathcal{L} = -\frac{1}{8} D^{\alpha\beta} D^*_{\alpha\beta}, \tag{3.9}$$

whereas the duality constraint (3.3) is

$$*D^{\alpha\beta} = -i D^{\alpha\beta}. \tag{3.10}$$

The Euler-Lagrange equations for the Lagrangian (3.9) with respect to $\mathrm{X}^\alpha$ or $\mathrm{X}^{*\alpha}$, or the Bianchi identity for the field (3.8), yield the Maxwell equations reduced now, by virtue of (3.10), to a single equation:

$$\partial_\beta D^{\alpha\beta} = 0. \tag{3.11}$$

The proper Hamiltonian of the field is readily recovered from the Lagrangian (3.9) using the Lagrangian formalism for complex fields (see [39], §II.A.2) and also the connection between the potentials and fields (3.6): $\partial_t \mathbf{X} = -i\nabla \times \mathbf{X} = -\mathbf{D}$. This yields $\mathcal{H} = \frac{1}{2}\mathbf{D}\cdot\mathbf{D}^* = \frac{1}{2}\left(E^2 + B^2\right) = W$.

The dual transformations (1.2) and (3.7) become a simple $U(1)$ gauge transformation in this complex formalism:

$$\mathrm{X}^\alpha \to e^{-i\theta}\mathrm{X}^\alpha, \quad D^{\alpha\beta} \to e^{-i\theta} D^{\alpha\beta}. \tag{3.12}$$

The Lagrangian (3.9) is obviously invariant with respect to (3.12). The corresponding conserved Noether current is easily obtained and equals:

$$J^{\alpha} = \frac{1}{2}\mathrm{Im}\left(D^{\alpha\beta}\mathrm{X}^{*}_{\beta}\right), \quad \partial_{\alpha}J^{\alpha} = 0. \tag{3.13}$$

This conservation law represents the *true helicity conservation* in Maxwell equations and coincides with equations (2.11) and (2.12) including the proper helicity and spin densities:

$$J^{0} \equiv H = \frac{1}{2}\left(\mathbf{A}\cdot\mathbf{B} - \mathbf{C}\cdot\mathbf{E}\right), \quad \mathbf{J} \equiv \mathbf{S} = \frac{1}{2}\left(\mathbf{E}\times\mathbf{A} + \mathbf{B}\times\mathbf{C}\right). \tag{3.14}$$

The same expressions for the helicity and spin densities were obtained in [8,12,16,27,28,35,43]. In the case of monochromatic fields, they are proportional to the components of the Lipkin's zilch current [22,23,29,33,35] and coincide with spin angular momentum in [42,44] (see Section 3.3 below). It is worth emphasizing that the derivation of the helicity current (3.13) and (3.14) from the standard Lagrangian (2.2) requires a sophisticated procedure described by Deser and Teitelboim [12], while with the dual-symmetric Lagrangian (3.2) or (3.9) it arises in a straightforward manner.

Furthermore, considering an infinitesimal time-translation transformation, $t \to t + \tau$, $\tau \to 0$, in a manner similar to (2.13)–(2.17), one can show that the Lagrangian (3.2) and (3.9) is transformed on the minimal-action trajectories (without taking into account the duality constraint) as:

$$\mathcal{L} \to \mathcal{L} - \tau\,\partial_{\alpha}\left[\frac{1}{2}\mathrm{Re}\left(D^{\alpha\beta}\partial_{t}\mathrm{X}^{*}_{\beta}\right)\right]. \tag{3.15}$$

In contrast to (2.13)–(2.15), this immediately evokes the *true* conserved energy-momentum current (2.16) and (2.17):

$$P^{\alpha} = -\frac{1}{2}\mathrm{Re}\left(D^{\alpha\beta}\partial_{t}\mathrm{X}^{*}_{\beta}\right), \quad \partial_{\alpha}P^{\alpha} = 0, \tag{3.16}$$

$$P^{0} \equiv W = \frac{1}{2}\left(E^{2} + B^{2}\right), \quad \mathbf{P} = \mathbf{E}\times\mathbf{B}. \tag{3.17}$$

Thus, our dual-symmetric Lagrangian formalism generates the same Maxwell equations, but also naturally reveals the proper helicity and energy conserved currents, containing the physically-meaningful dual-symmetric helicity, spin, and energy densities. Below we provide a complete Noether analysis of conservation laws associated with the Poincaré symmetries and show that the choice of dual-symmetric Lagrangian (3.2) or (3.9) makes an important difference in the canonical form of the conservation laws.

### *3.2. Energy-momentum and angular momentum tensors.*

In this Section, we follow the plan of Section 2.2, but now using the dual-symmetric Lagrangian (3.2) and its representation (3.9) via complex Riemann-Silberstein fields $\mathrm{X}^{\alpha} = A^{\alpha} + iC^{\alpha}$ and $D^{\alpha\beta} = F^{\alpha\beta} + iG^{\alpha\beta}$. We will consider canonical and symmetrized energy-momentum and angular-momentum Noether currents and compare them with their counterparts obtained previously within standard electromagnetism.

*3.2.1. Canonical tensors.* First, the canonical energy-momentum tensor following from Lagrangian (3.9) is:

$$T^{\alpha\beta} = \frac{1}{2}\mathrm{Re}\left[\left(\partial^{\alpha}\mathrm{X}^{\gamma}\right)D^{*\beta}{}_{\gamma}\right], \quad \partial_{\beta}T^{\alpha\beta} = 0. \tag{3.18}$$

This tensor is non-symmetric, $T^{\alpha\beta} \neq T^{\beta\alpha}$, and the corresponding 4-momentum density is given by $P_{O}^{\alpha} = T^{\alpha 0} = \frac{1}{2}\mathrm{Re}\left[\mathbf{R}^{*}\cdot\left(\partial^{\alpha}\right)\mathbf{X}\right]$. This yields the proper energy density $P_{O}^{0} = W = \frac{1}{2}\left(E^{2} + B^{2}\right)$, and also the following *orbital* energy flux density, the dual-symmetric modification of (2.20):

$$\mathbf{P}_O = \frac{1}{2}\left[\mathbf{E}\cdot(\nabla)\mathbf{A} + \mathbf{B}\cdot(\nabla)\mathbf{C}\right]. \tag{3.19}$$

In turn, the total energy flux density, is given by $P^\alpha = T^{0\alpha} = (W, \mathbf{P})$, in agreement with (3.16) and (3.17).

Second, akin to Eq. (2.21), the canonical angular momentum tensor takes on the form

$$M^{\alpha\beta\gamma} = r^\alpha T^{\beta\gamma} - r^\beta T^{\alpha\gamma} + S^{\alpha\beta\gamma}, \quad \partial_\gamma M^{\alpha\beta\gamma} = 0, \tag{3.20}$$

where the spin tensor is

$$S^{\alpha\beta\gamma} = \frac{1}{2}\mathrm{Re}\left(\mathrm{X}^\beta D^{*\gamma\alpha} - \mathrm{X}^\alpha D^{*\gamma\beta}\right), \quad \partial_\gamma S^{\alpha\beta\gamma} = T^{\alpha\beta} - T^{\beta\alpha} \neq 0. \tag{3.21}$$

Calculating the pseudo-vector of the angular momentum density, $M_i = \frac{1}{2}\varepsilon_{ijk} M^{jk0}$, we arrive at the following orbital and spin parts, $\mathbf{M} = \mathbf{L} + \mathbf{S}$ [cf. (2.23)]:

$$\mathbf{L} = \frac{1}{2}\left[\mathbf{E}\cdot(\mathbf{r}\times\nabla)\mathbf{A} + \mathbf{B}\cdot(\mathbf{r}\times\nabla)\mathbf{C}\right] = \mathbf{r}\times\mathbf{P}_O, \tag{3.22}$$

$$\mathbf{S} = \frac{1}{2}\left(\mathbf{E}\times\mathbf{A} + \mathbf{B}\times\mathbf{C}\right). \tag{3.23}$$

In turn, the boost-momentum components in the tensor (3.20), $N_i = M^{0i0}$, yield [cf. (2.24) and (2.25)]:

$$\mathbf{N} = \mathbf{N}_O = \frac{1}{2}\left[\mathbf{E}\cdot(t\nabla + \mathbf{r}\partial_t)\mathbf{A} + \mathbf{B}\cdot(t\nabla + \mathbf{r}\partial_t)\mathbf{C}\right] = t\,\mathbf{P}_O - \mathbf{r}W, \quad \mathbf{N}_S = 0. \tag{3.24}$$

Thus, the components (3.19), (3.22)−(3.24) of the canonical energy-momentum and angular-momentum tensors contain now the proper conserved energy density $W$, spin density $\mathbf{S}$ coinciding with the conserved helicity flux density (2.11) or (3.14), and the corresponding orbital momentum density $\mathbf{P}_O$. Therefore, *the dual electromagnetism fixes all the discrepancies in the canonical Noether currents of the standard electromagnetic theory*. This approach naturally contains a meaningful dual-symmetric separation of the spin and orbital degrees of freedom [3,42,43], consistent with the helicity and energy conservation laws (see also Fig. 1).

*3.2.2. Symmetrized tensors.* The Belinfante's symmetrization procedure can be applied to the canonical energy-momentum tensor (3.18), which results in the same symmetric energy-momentum tensor $\mathcal{T}^{\alpha\beta}$ as in Eqs. (2.26) (but now simplified in the dual-complex form):

$$\mathcal{T}^{\alpha\beta} = T^{\alpha\beta} + \partial_\gamma K^{\alpha\beta\gamma} = \frac{1}{2} D^{\alpha\gamma} D^{*\beta}{}_{\gamma}, \tag{3.25}$$

where $K^{\alpha\beta\gamma} = -\frac{1}{2}\mathrm{Re}\left(\mathrm{X}^\alpha D^{*\beta\gamma}\right)$. The components of $\mathcal{T}^{\alpha\beta}$ are displayed in Eqs. (2.28) and (2.29). The spin current is obtained as $P_{Si} = \partial_\gamma K^{i0\gamma}$, which yields the dual-symmetric version of Eq. (2.30):

$$\mathbf{P}_S = -\frac{1}{2}\left[(\mathbf{E}\cdot\nabla)\mathbf{A} + (\mathbf{B}\cdot\nabla)\mathbf{C}\right]. \tag{3.26}$$

Akin to Eqs. (2.31)−(2.33), the Poynting vector is the sum of the orbital and spin energy flux densities (3.19) and (3.26):

$$\mathbf{P} = \mathbf{P}_O + \mathbf{P}_S, \tag{3.27}$$

where the spin current makes no contribution to the integral momentum:

$$\int \mathbf{P}_S\, dV = 0. \tag{3.28}$$

The circulations of the orbital and spin currents produce, respectively, the orbital and spin angular momenta (3.22) and (3.23):

$$\mathbf{L} = \mathbf{r}\times\mathbf{P}_O, \quad \int \mathbf{S}\, dV = \int \mathbf{r}\times\mathbf{P}_S\, dV. \tag{3.29}$$

The dual-symmetric spin and orbital parts of the energy flux density (3.19) and (3.26) coincide with those suggested recently in [3,42,51].

Evidently, the symmetrized angular-momentum tensor also coincides in the standard- and dual-electromagnetism approaches:

$$\mathcal{M}^{\alpha\beta\gamma} = r^{\alpha}\mathcal{T}^{\beta\gamma} - r^{\beta}\mathcal{T}^{\alpha\gamma} . \tag{3.30}$$

Its components are described by Eqs. (2.35). This suggests that the symmetrized energy-mometum and angular-momentum tensors are independent of the choice of Lagrangian. The price of this independence is the impossibility to separate the spin and orbital degrees of freedom and to trace the connection with the helicity conservation.

*3.2.3. Integral conserved quantities.* As usual, differential conservation laws can be written in the form of integral conserved quantities. In dual electromagnetism, they are similar to their standard counterparts, and, furthermore, a number of them coincide in the two formalisms. In a way entirely analogous to the derivation of Eqs. (2.38)−(2.42), we obtain:

$$\int T^{\alpha 0}\, dV = \int \tilde{T}^{\alpha 0}\, dV = \int \mathcal{T}^{\alpha 0}\, dV = \text{const}, \tag{3.31}$$

$$\int W dV = \text{const}, \quad \int \mathbf{P}_O\, dV = \int \tilde{\mathbf{P}}_O\, dV = \int \mathbf{P}\, dV = \text{const}. \tag{3.32}$$

$$\int M^{\alpha\beta 0}\, dV = \int \tilde{M}^{\alpha\beta 0}\, dV = \int \mathcal{M}^{\alpha\beta 0}\, dV = \text{const}, \tag{3.33}$$

$$\int \left(\mathbf{r}\times\mathbf{P}_O + \mathbf{S}\right) dV = \int \left(\mathbf{r}\times\tilde{\mathbf{P}}_O + \tilde{\mathbf{S}}\right) dV = \int \left(\mathbf{r}\times\mathbf{P}\right) dV = \text{const}, \tag{3.34}$$

$$\int \left(t\,\mathbf{P}_O - \mathbf{r}W\right) dV = \int \left(t\,\tilde{\mathbf{P}}_O - \mathbf{r}\tilde{W}\right) dV = \int \left(t\,\mathbf{P} - \mathbf{r}W\right) dV = \text{const}. \tag{3.35}$$

Note that equation (2.44) for the rectilinear motion of the energy centroid immediately follows from the canonical dual-symmetric boost momentum (3.24) $\mathbf{N}$ and Eq. (3.32), in contrast to the canonical (2.25) $\tilde{\mathbf{N}}$ in the standard approach.

At the same time, the integral spin and orbital angular momenta are *different* in the standard and dual approaches, although they are all conserved quantities [see (2.46) and (2.47)]:

$$\text{const} = \int \mathbf{L}\, dV \neq \int \tilde{\mathbf{L}}\, dV = \text{const}, \quad \text{const} = \int \mathbf{S}\, dV \neq \int \tilde{\mathbf{S}}\, dV = \text{const}. \tag{3.36}$$

And, again, for the sake of completeness, we repeat here Eqs. (2.36) and (2.37) involving integral forms of the true and false helicity and energy that appear in the standard and dual formalisms:

$$\text{const} = \int H\, dV \neq \int \tilde{H}\, dV \neq \text{const}, \quad \text{const} = \int W dV \neq \int \tilde{W} dV \neq \text{const}. \tag{3.37}$$

From a complete set of equations (3.31)−(3.37), augmented by relations (2.31), (2.33), (3.27), and (3.29), one can see that the integral conserved quantities of the dual electromagnetism (i.e., those without "tilde") form a perfectly consistent system, as opposed to those of the standard electromagnetic theory. In Section 3.3 we show that the dual-symmetric approach is also in agreement with the quantum-like operator formalism [3,42,44].

### *3.3. Monochromatic fields and operator representation.*

In a vast majority of optical problems, monochromatic electromagnetic fields and their time-averaged characteristics are considered. In this case, all linear field characteristics ($\mathbf{O} = \mathbf{A}, \mathbf{C}, \mathbf{E}, \mathbf{B}$) have the form $\mathbf{O}(\mathbf{r},t) = \text{Re}\left[\mathbf{O}(\mathbf{r})e^{-i\omega t}\right]$, where $\omega$ is the frequency and $\mathbf{O}(\mathbf{r})$ is the complex field amplitude. Substituting this into Eqs. (3.6), we find that the complex amplitudes of the potentials and fields become proportional to each other:

$$i\omega\mathbf{A} = \mathbf{E}, \quad i\omega\mathbf{C} = \mathbf{B}. \tag{3.38}$$

Next, the time-average (over one period of oscillations) of any real quadratic field form $F = \mathbf{O}\,\hat{f}\,\mathbf{O}$ becomes

$$\bar{F} = \frac{1}{2}\text{Re}\left(\mathbf{O}^{*}\hat{f}\,\mathbf{O}\right). \tag{3.39}$$

The dynamical characteristics of the field (helicity, energy, momentum, etc.) considered in previous sections represent quadratic forms with respect to the fields and potentials. Then, applying the time averaging (3.39) with relations (3.38) to the densities of energy-momentum (2.28), helicity (2.11) or (3.14), spin and orbital momenta (3.19), (3.26) and angular momenta (3.22), (3.23), and boost momentum (3.24), we obtain

$$\bar{W} = \frac{1}{4}\left(|\mathbf{E}|^2 + |\mathbf{B}|^2\right), \quad \bar{\mathbf{P}} = \frac{1}{2}\mathrm{Re}\left(\mathbf{E}^* \times \mathbf{B}\right), \tag{3.40}$$

$$\bar{H} = -\frac{1}{2\omega}\mathrm{Im}\left(\mathbf{E}^* \cdot \mathbf{B}\right), \tag{3.41}$$

$$\bar{\mathbf{P}}_O = \frac{1}{4\omega}\mathrm{Im}\left[\mathbf{E}^* \cdot (\nabla)\mathbf{E} + \mathbf{B}^* \cdot (\nabla)\mathbf{B}\right], \tag{3.42}$$

$$\bar{\mathbf{P}}_S = \frac{1}{8\omega}\nabla \times \mathrm{Im}\left[\mathbf{E}^* \times \mathbf{E} + \mathbf{B}^* \times \mathbf{B}\right], \tag{3.43}$$

$$\bar{\mathbf{L}} = \frac{1}{4\omega}\mathrm{Im}\left[\mathbf{E}^* \cdot (\mathbf{r} \times \nabla)\mathbf{E} + \mathbf{B}^* \cdot (\mathbf{r} \times \nabla)\mathbf{B}\right], \tag{3.44}$$

$$\bar{\mathbf{S}} = \frac{1}{4\omega}\mathrm{Im}\left(\mathbf{E}^* \times \mathbf{E} + \mathbf{B}^* \times \mathbf{B}\right), \tag{3.45}$$

When deriving (3.43) we used Maxwell equations $\nabla \cdot \mathbf{E} = \nabla \cdot \mathbf{B} = 0$. Note that the time-averaged boost momentum $\bar{\mathbf{N}}$ does not give a meaningful result for the monochromatic field, because $\mathbf{N}$ in Eq. (3.24) explicitly contains time $t$. In addition, a monochromatic field cannot be localized in three dimensions, and the energy centroid (related to the boost-momentum conservation) becomes ill-defined. At the same time, the averaged total angular momentum density (2.35) from the symmetrized angular-momentum tensor $\mathcal{M}^{\alpha\beta\gamma}$ is naturally expressed via the energy and momentum (3.40):

$$\bar{\boldsymbol{\mathcal{M}}} = \mathbf{r} \times \bar{\mathbf{P}}. \tag{3.46}$$

The dual-symmetric and gauge-invariant expressions (3.40)–(3.45) coincide with results obtained in different contexts in recent works [3,33,42,44,51,53]. In particular, Eqs. (3.41) and (3.45) demonstrate the proportionality of the helicity and spin densities to the Lipkin's zilch pseudo-scalar and pseudo-vector [33–35]. Remarkably, for a monochromatic field, the time averages of the dual-asymmetric helicity (2.9) and the true helicity (2.11) or (3.14) coincide with each other:

$$\bar{H} = \bar{\tilde{H}}. \tag{3.47}$$

At the same time, all other quantities – energy, spin and orbital energy fluxes and angular momenta – remain essentially different in their dual-symmetric and asymmetric 'tilded' versions:

$$\bar{W} \neq \bar{\tilde{W}}, \quad \bar{\mathbf{P}}_O \neq \bar{\tilde{\mathbf{P}}}_O, \quad \bar{\mathbf{P}}_S \neq \bar{\tilde{\mathbf{P}}}_S, \quad \bar{\mathbf{L}} \neq \bar{\tilde{\mathbf{L}}}, \quad \bar{\mathbf{S}} \neq \bar{\tilde{\mathbf{S}}}. \tag{3.48}$$

Equation (3.47) explains why the false dual-asymmetric helicity $\tilde{H}$ is associated with the dual-symmetric conserved Lipkin's zilch in [34]: Monochromatic fields and integral quantities are considered there. Rigorously speaking, the integral characteristics diverge for monochromatic fields, as it cannot be localized and contains infinite number of photons. Note also that the spin energy flux density (3.43) represents the curl of the spin density (3.45): $\bar{\mathbf{P}}_S = \frac{1}{2}\nabla \times \bar{\mathbf{S}}$. This reveals the divergence-less character of the spin current. According to (3.28) and (3.29) this current makes no contribution to the integral momentum of the field, and only generates a purely intrinsic spin angular momentum [42,48–53]. An entirely similar spin current generates the spin of the relativistic quantum electron [48–50,57].

Importantly, equations (3.40)–(3.45) reveal profound quantum-mechanical analogies and can be reduced to the following simple forms:

$$\bar{W} = (\psi \,|\, \hat{w} \,|\, \psi), \tag{3.40'}$$

$$\bar{H} = (\psi \,|\, \frac{\hat{\mathbf{p}}\cdot\hat{\mathbf{S}}}{p} \,|\, \psi), \tag{3.41'}$$

$$\bar{\mathbf{P}}_O = (\psi \,|\, \hat{\mathbf{p}} \,|\, \psi), \tag{3.42'}$$

$$\bar{\mathbf{P}}_S = (\psi \,|\, i\,\hat{\mathbf{p}} \times \hat{\mathbf{S}} \,|\, \psi), \tag{3.43'}$$

$$\bar{\mathbf{L}} = (\psi \,|\, \hat{\mathbf{r}} \times \hat{\mathbf{p}} \,|\, \psi), \tag{3.44'}$$

$$\bar{\mathbf{S}} = (\psi \,|\, \hat{\mathbf{S}} \,|\, \psi). \tag{3.45'}$$

Here we used the *dual-symmetric state vector* $|\psi) = \frac{1}{2\sqrt{\omega}}\left|\mathbf{E},\mathbf{B}\right)$, the inner product assumes the real part of the scalar product (without volume integration, since here we calculate local densities), whereas the operators are:

$$\hat{w} = \omega\,, \quad \hat{\mathbf{p}} = -i\nabla\,, \quad \hat{\mathbf{r}} = \mathbf{r}\,, \quad \left(\hat{S}_a\right)_{ij} = -i\varepsilon_{ija}\,. \tag{3.49}$$

The spin-1 matrix operator $\hat{\mathbf{S}}$ acts as $\mathbf{O}^*\cdot\left(\hat{\mathbf{S}}\right)\mathbf{O} = \mathrm{Im}\left[\mathbf{O}^*\times\mathbf{O}\right]$, and we also used Maxwell equations and the dispersion relation $p^2 = \omega^2$. Importantly, within the standard electromagnetism of Section 2, the dual-asymmetric 'tilded' quantities (3.40')–(3.45') would be given by the same equations, with the same operators, but with the *dual-asymmetric state vector* $|\tilde{\psi}) = \frac{1}{\sqrt{2\omega}}\left|\mathbf{E}\right)$. But some of these quantities (e.g., the false energy $\tilde{W}$) are not conserved, and even the standard electromagnetism operates with the proper energy $W$ which require the use of the dual-symmetric state vector $|\psi) = \frac{1}{2\sqrt{\omega}}\left|\mathbf{E},\mathbf{B}\right)$ (see, e.g., [44,45], where the Riemann-Silberstein vector is considered as a natural choice for the photon wave function). Therefore, *within the standard electromagnetism, it is impossible to write all characteristics of the field in a consistent quantum-like operator form (3.40')–(3.45')*. Here we do *not* consider quantization of the fields, but even the first-quantization formalism for classical fields shows that only the dual-symmetric formulation of electromagnetism yields meaningful and mutually-consistent expectation values of quantum spin and orbital operators with the suitable state vector.

Equations (3.40')–(3.45') allow a straightforward transition to the Fourier (momentum) representation [23,27,28,33,42,44,51,52]: $\left|\psi(\mathbf{r})\right) \to \left|\psi(\mathbf{k})\right)$. In doing so, one merely has to change the momentum and coordinate operators (3.49) as $\hat{\mathbf{p}} = \mathbf{k}$ and $\hat{\mathbf{r}} = i\nabla_{\mathbf{k}}$. In the Fourier representation, it becomes clear that the helicity represents the difference in number of right-hand and left-hand circularly-polarized plane waves [23,27,28,33–35]. If the field has a well-defined quantum helicity $\sigma = \pm 1$ (i.e., the Fourier spectrum of the field contains only plane waves with one circular polarization $\sigma$), then the complex electric and magnetic amplitudes are related as $\mathbf{E} = i\sigma\mathbf{B}$. In this case, the helicity (3.41), spin (3.45), energy and momentum (3.40) become simply related as:

$$\bar{H} = \sigma\frac{\bar{W}}{\omega}\,, \quad \bar{\mathbf{S}} = \sigma\frac{\bar{\mathbf{P}}}{\omega}\,. \tag{3.50}$$

Then, the helicity conservation (2.11), (2.12) or (3.13), (3.14) becomes equivalent to the energy conservation (Poynting theorem) (2.16), (2.17) or (3.16), (3.17) [33]. Furthermore, in such pure helicity state, the electric and magnetic contributions are equal in all quantities (3.40)–(3.46), so that the standard calculations of Section 2 and the dual-symmetric formalism are equivalent. However, for a generic field, containing different helicity states in the Fourier spectrum, the helicity conservation represents a truly independent conservation law, whereas the dual-symmetric quantities (3.40), (3.42)–(3.45) differ from their 'tilded' counterparts obtained within the standard approach.

<table>
<tr><th colspan="2"></th><th>Standard</th><th>Dual</th></tr>
<tr><td colspan="2">Lagrangian</td><td>$\tilde{\mathcal{L}} = -\frac{1}{4} F^{\alpha\beta} F_{\alpha\beta}$</td><td>$\mathcal{L} = -\frac{1}{8} D^{\alpha\beta} D^{*}_{\alpha\beta}$</td></tr>
<tr><td colspan="2">Dual current</td><td colspan="2">$J^{\alpha} = \frac{1}{2} \mathrm{Im}\left(D^{\alpha\beta} \mathrm{X}^{*}_{\beta}\right) = \frac{1}{2}(H, \mathbf{S})$</td></tr>
<tr><td rowspan="4">Poincaré currents: energy-momentum and angular momentum</td><td rowspan="2">Canonical</td><td>$\tilde{T}^{\alpha\beta} = F^{\beta\gamma} \partial^{\alpha} A_{\gamma} + g^{\alpha\beta} \tilde{\mathcal{L}}$<br>$\tilde{T}^{\alpha 0} = (W, \tilde{\mathbf{P}}_O)$</td><td>$T^{\alpha\beta} = \mathrm{Re}\left(D^{*\beta\gamma} \partial^{\alpha} \mathrm{X}_{\gamma}\right)/2$<br>$T^{\alpha 0} = (W, \mathbf{P}_O)$</td></tr>
<tr><td>$\tilde{M}^{\alpha\beta\gamma} = x^{\alpha} \tilde{T}^{\beta\gamma} - x^{\beta} \tilde{T}^{\alpha\gamma} + \tilde{S}^{\alpha\beta\gamma}$<br>$\tilde{M}^{\alpha\beta 0} = \begin{pmatrix} \tilde{\mathbf{N}} = t \tilde{\mathbf{P}}_O - \mathbf{r} \tilde{W} \\ \tilde{\mathbf{M}} = \mathbf{r} \times \tilde{\mathbf{P}}_O + \tilde{\mathbf{S}} \end{pmatrix}$</td><td>$M^{\alpha\beta\gamma} = x^{\alpha} T^{\beta\gamma} - x^{\beta} T^{\alpha\gamma} + S^{\alpha\beta\gamma}$<br>$M^{\alpha\beta 0} = \begin{pmatrix} \mathbf{N} = t \mathbf{P}_O - \mathbf{r} W \\ \mathbf{M} = \mathbf{r} \times \mathbf{P}_O + \mathbf{S} \end{pmatrix}$</td></tr>
<tr><td rowspan="2">Symmetrized</td><td colspan="2">$\mathcal{T}^{\alpha\beta} = F^{\alpha\gamma} F^{\beta}{}_{\gamma} + g^{\alpha\beta} \tilde{\mathcal{L}} = \frac{1}{2} D^{\alpha\gamma} D^{*\beta}{}_{\gamma}, \quad \mathcal{T}^{\alpha 0} = (W, \mathbf{P})$</td></tr>
<tr><td colspan="2">$\mathcal{M}^{\alpha\beta\gamma} = x^{\alpha} \mathcal{T}^{\beta\gamma} - x^{\beta} \mathcal{T}^{\alpha\gamma} = \begin{pmatrix} \mathbf{N} = t \mathbf{P} - \mathbf{r} W \\ \mathbf{M} = \mathbf{r} \times \mathbf{P} \end{pmatrix}$</td></tr>
</table>

**Figure 1.** Summary of the main Noether currents which appear in the standard (Section 2) and dual (Section 3) versions of electromagnetism. The helicity current $J^{\alpha}$ and symmetrized Poincaré currents (energy-momentum and angular momentum tensors, $\mathcal{T}^{\alpha\beta}$ and $\mathcal{M}^{\alpha\beta\gamma}$) are common and dual-symmetric. At the same time, the canonical energy-momentum and angular-momentum tensors, $T^{\alpha\beta}$ and $M^{\alpha\beta\gamma}$, conflict with the helicity and energy conservation laws in the standard electromagnetic theory. These discrepancies disappear in the dual electromagnetism. Inconsistent quantities are shown in red, as opposed to the consistent ones, shown in green. The fields are characterized by the magnetic four-potential $A^{\alpha}$, field tensor $F^{\alpha\beta}$, their dual counterparts: electric four-potential $C^{\alpha}$ and dual field tensor $G^{\alpha\beta}$, and combined complex quantities $\mathrm{X}^{\alpha} = A^{\alpha} + i C^{\alpha}$ and $D^{\alpha\beta} = F^{\alpha\beta} + i G^{\alpha\beta}$.

## 4. Observability and relation to quantum weak measurements

Here we briefly discuss observable consequences of the dual symmetry and quantities that make a difference between the standard and dual formulations of electromagnetism. It is important to emphasize that although we discuss properties of the *free* electromagnetic field, they are experimentally observed only via various *light-matter interactions* (any detector involves matter).

First, we note that the dual symmetry and *conservation of helicity* $H$ are not abstract properties, but they have immediately observable consequences. In particular, it follows that any perturbation which does *not* break the electric-magnetic symmetry will keep the electromagnetic helicity as an exact invariant of the problem. This explains the conservation of the helicity of photons in an *arbitrary* gravitational field [45]. Furthermore, the helicity turns out to be exactly conserved in any optical scattering on macroscopic objects with equal electric and magnetic constants $\varepsilon = \mu$ [36,45,58], which is usually interpreted as matching of optical impedances $\varepsilon / \mu = \mathrm{const}$. At the same time, any coupling with matter having *asymmetric* electric and magnetic properties will in general produce a conversion between the two helicity states of photons. As the helicity conservation involves the spin angular momentum, the conversion of the helicity is usually accompanied by a conversion between the spin and orbital angular momenta [36,59], although the opposite is not generally true. Thus, the dual symmetry and helicity conservation offer an additional integral of motion which can be used in the analysis of various optical interactions.

Second, let us consider physical quantities which appear to be essentially different in the standard and dual versions of electromagnetism – the *spin and orbital angular momenta of light*, $\mathbf{S}$ and $\mathbf{L}$. Although quantum electrodynamics sometimes concerns the separation of the spin and orbital parts of the photon angular momentum as physically meaningless, modern optics points to the independent observability of these quantities [41,54,55]. Indeed, locally the orbital and spin parts of the angular momenta of light cause qualitatively different motions (orbiting and spinning) of probe particles immersed in the field [54]. In quantum electrodynamics, the photon interaction with an atom also causes changes in the extrinsic and intrinsic angular momentum of the atom, quite similar to the orbital and spinning motion of the classical probe particle [41]. In addition, the integral value of the intrinsic orbital angular momentum is closely related to the spatial distribution of the field intensity and its localizability [42,60]. Thus, if we regard the spin and orbital angular momenta of the electromagnetic field as separately measurable quantities (either in their local or integral values), this allows to discriminate between the standard and dual electromagnetic theories.

The spin and orbital angular momenta of the field are generated, respectively, by *spin and orbital* energy fluxes $\mathbf{P}_S$ and $\mathbf{P}_O$ [3,37,42,48–53], see Eqs. (2.33) and (3.29), which together form the Poynting vector $\mathbf{P} = \mathbf{P}_O + \mathbf{P}_S$. These local quantities are *different* in the standard and dual theories. Remarkably, although the Poynting vector $\mathbf{P}$ is usually considered as a physically-meaningful quantity, it turns out that the local orbital energy flux $\mathbf{P}_O$ (i.e., the *canonical* momentum density) can be measured easier and in a more straightforward way via the motion of a probe particle [51,54,56]. Indeed, it is the orbital energy flux that transports energy, represents the local expectation value of the momentum operator [see Eqs. (3.42) and (3.42')], and it can be associated with the standard quantum-mechanical probability current [3,48–50]. (In contrast, the spin energy flux, $\mathbf{P}_S$, is sometimes regarded as a virtual divergence-less current which cannot be observed *per se* [48–50].) Thus, measuring the canonical momentum density $\mathbf{P}_O$, one can also discriminate between the standard and dual theories.

The observability of the *local* densities and currents is an important problem by itself. In standard field theories, all local densities are usually interpreted as unobservable auxiliary quantities, whereas only the integral energy, momentum, and angular momentum of the field make physical sense. However, classical optics naturally regard the local energy density $W$, momentum density $\mathbf{P}$, and other currents, as meaningful and observable characteristics of the field [3,51]. Moreover, it seems that quantum measurements also allow the detection of local currents [3,61]. This is related to the concept of *quantum weak measurements* [62–64]. Both classical-optics and quantum weak-measurement approaches are based here on a natural idea: The straightforward way to measure a current of any flow is to place a small probe particle in the flow and to trace its motion.

In classical optical fields, a small particle experiences the action of the radiation pressure force and moves proportional to the local momentum density of the field [54,56]. For small Rayleigh particles this force is proportional to the *canonical* (*orbital*) momentum density $\mathbf{P}_O$ rather than to the Poynting vector $\mathbf{P}$ [3,56]. Furthermore, the particle spins proportionally to the local spin density of the field [54]. Thus, by measuring the velocity of the linear motion of the particle and the angular velocity of its spinning motion, one can determine the local momentum density $\mathbf{P}_O$ (hence, also the orbital angular momentum density $\mathbf{L} = \mathbf{r} \times \mathbf{P}_O$) and the spin density $\mathbf{S}$.

The same experiments can be interpreted within the quantum weak-measurement picture [3,61–64]. Let us represent the field distribution as a quantum photon state $|\psi(\mathbf{r})\rangle$ with indeterminate coordinate (i.e., the spread is much larger than the size of the probe particle). If this photon interacts with the particle located at $\mathbf{r}=\mathbf{r}_0$, this fixes the coordinate of the photon, i.e., the particle *post-selects* the photon in the state with well-defined coordinate, $|\mathbf{r}_0\rangle$. Although the photon-particle interaction has a very low probability $\langle \mathbf{r}_0|\psi\rangle \ll 1$, averaging over many events (as it happens with classical multi-photon fields) provides *simultaneous* information about the *position* of the photon and its *local momentum* (current). This is expressed via a the following quantum weak-measurement equation [3,61]:

$$\langle \mathbf{p}\rangle_{weak} = \mathrm{Re}\frac{\langle \mathbf{r}_0|\hat{\mathbf{p}}|\psi\rangle}{\langle \mathbf{r}_0|\psi\rangle} = \mathrm{Re}\frac{\langle \psi|\mathbf{r}_0\rangle\langle \mathbf{r}_0|\hat{\mathbf{p}}|\psi\rangle}{\langle \psi|\mathbf{r}_0\rangle\langle \mathbf{r}_0|\psi\rangle} = \frac{\mathbf{j}(\mathbf{r}_0)}{\rho(\mathbf{r}_0)}. \tag{4.1}$$

Here $\rho(\mathbf{r}) = (\psi(\mathbf{r})|\psi(\mathbf{r}))$ and $\mathbf{j}(\mathbf{r}) = (\psi(\mathbf{r})|\hat{\mathbf{p}}|\psi(\mathbf{r}))$ are the local probability density and current in the field. As described in Section 3.3, we have $\rho = \overline{W}/\omega$ and $\mathbf{j} = \overline{\mathbf{P}}_O$ for a monochromatic electromagnetic field, so that the quantum weak measurement of the field momentum is essentially the local measurements of the orbital current $\mathbf{P}_O$. Recently, the same weak-measurement scheme (4.1) (but employing another, non-particle, detector) was successfully used to detect the local photon 'trajectories' in the double-slit experiment [65]. These Bohmian trajectories are nothing but the streamlines of the orbital energy flux $\mathbf{P}_O$. Furthermore, the momentum exchange in the resonant interaction between a moving atom and an electromagnetic wave also reveals the local value of the canonical momentum. This is seen, e.g., in the Doppler-shift experiments with evanescent waves, where $\langle \mathbf{p}\rangle_{weak} > \omega$ [66], in agreement with a 'superluminal' character of the orbital energy flux discussed in [53]. For the local measurements of the spin angular momentum density via a spinning particle, one can write a weak-measurement equation similar to (4.1):

$$\langle \mathbf{S}\rangle_{weak} = \mathrm{Re}\frac{\langle \mathbf{r}_0|\hat{\mathbf{S}}|\psi\rangle}{\langle \mathbf{r}_0|\psi\rangle} = \mathrm{Re}\frac{\langle \psi|\mathbf{r}_0\rangle\langle \mathbf{r}_0|\hat{\mathbf{S}}|\psi\rangle}{\langle \psi|\mathbf{r}_0\rangle\langle \mathbf{r}_0|\psi\rangle} = \frac{\mathbf{S}(\mathbf{r}_0)}{\rho(\mathbf{r}_0)}. \tag{4.2}$$

It might seem that the above local measurements of optical currents and angular momenta solve the problem and result in unambiguous and objective determination of the field properties. However, this is not so. The problem is that the results of the measurements based on a probe particle (i.e., involving *light-matter interaction*) crucially depend on the *properties of the particle*. For instance, the light scattering on a small *dielectric* particle can be considered in the *electric*-dipole (Rayleigh) approximation, and then it turns out that the radiation force that pushes the particle is proportional to the *electric* orbital momentum density [3,56] [cf. Eq. (3.42)]:

$$\mathbf{F}_{\mathrm{rad}} \propto \overline{\mathbf{P}}_O^{\mathrm{electric}} = \frac{1}{2\omega}\mathrm{Im}\left[\mathbf{E}^*\cdot(\nabla)\mathbf{E}\right]. \tag{4.3}$$

This is clearly a dual-*asymmetric* expression consistent with the *standard* electromagnetism and momentum density $\tilde{\mathbf{P}}_O$ defined in Eq. (2.20). However, if the same measurement is made by a small *magnetic* particle [56,67], and the interaction has a *magnetic*-dipole character, the radiation force will be proportional to a similar *magnetic* expression for the orbital momentum density:

$$\mathbf{F}_{\mathrm{rad}} \propto \overline{\mathbf{P}}_O^{\mathrm{magnetic}} = \frac{1}{2\omega}\mathrm{Im}\left[\mathbf{B}^*\cdot(\nabla)\mathbf{B}\right]. \tag{4.4}$$

Finally, only a particle with equivalent *electric and magnetic* properties will measure the *dual-symmetric* orbital momentum density $\mathbf{P}_O$, Eqs. (3.19) and (3.42):

$$\mathbf{F}_{\mathrm{rad}} \propto \overline{\mathbf{P}}_O = \frac{1}{4\omega}\mathrm{Im}\left[\mathbf{E}^*\cdot(\nabla)\mathbf{E} + \mathbf{B}^*\cdot(\nabla)\mathbf{B}\right]. \tag{4.5}$$

Similar observations can be made for measurements of the spin and orbital angular momenta. For instance, in paper [41] the authors point to the separate observability of the spin and orbital angular momenta of light, and analyse the photon interaction with an atom. This interaction is approximated by the *electric*-dipole coupling, and, due to this, the changes in the atomic states would measure the dual-*asymmetric* angular momenta $\tilde{\mathbf{L}}$ and $\tilde{\mathbf{S}}$ following from the standard approach.

Hence, the results of measurements of the dynamical characteristics of the electromagnetic field depend critically on the properties of the measuring device. They can naturally be dual-asymmetric as a consequence of the electric-magnetic asymmetry in matter (absence of magnetic charges). However, this does *not* mean that we should ascribe dual-asymmetric features to the *free* electromagnetic field. Indeed, in practice, it is difficult even to measure the energy density $\bar{W} = \left(|\mathbf{E}|^2 + |\mathbf{B}|^2\right)/4$ or Poynting vector $\bar{\mathbf{P}} = \frac{1}{2}\mathrm{Re}\left(\mathbf{E}^* \times \mathbf{B}\right)$, which are natural dual-symmetric conserved characteristics of the field. Typically, only the *electric* energy density, i.e., $\bar{\tilde{W}} = |\mathbf{E}|^2/2$, can be measured. Obviously, such asymmetry of measuring interactions does not suggest that we should associate the *non-conserved* dual-asymmetric quantity $\tilde{W} = E^2$ with the energy density of the electromagnetic field! Thus, in spite of such difficulties with measurements, if we would like to ascribe fundamental dynamical characteristics to the electromagnetic field *per se*, we have to maintain the electric-magnetic symmetry which is inherent in the free field. Naturally, only the dual electromagnetic theory suggested here provides such characteristics of the fields. Formally, they can be thought as a result of measurements made by an electromagnetically-neutral (e.g., gravitational or macroscopic with $\varepsilon = \mu$) detector.

To conclude this Section and support our arguments, let us mention two recent examples where an improper dual-asymmetric interpretation of the field and measurement properties brought about confusing results.

First, Tang and Cohen introduced a novel concept of "superchiral light", i.e., light that shows optical chirality (helicity) higher than that of a circularly polarized plane wave [29,30]. They observed that the so-called dissymmetry factor in local light interaction with a chiral particle can be anomalously large in such "superchiral" field configurations. In fact, it was shown in [33,34] that this enhanced chiral response arises *not* from extraordinary properties of the optical chirality or helicity density – it can never exceed the limit $\left|\bar{H}/\bar{W}\right|_{\max} = 1/\omega$ of a circularly-polarized plane wave. Instead, this is a property of the *particle* involving the *electric*-dipole interaction which is sensitive only to the *electric* energy density, i.e., $\bar{\tilde{W}} = |\mathbf{E}|^2/2$. This electric energy density appears in the denominator of the dissymmetry factor and causes its enhancement in the vicinity of the electric-field nodes. If a magnetic-dipole coupling of equal strength would also be present, the dissymmetry factor would have never been larger than that of a circularly-polarized plane wave [68].

Second, some of us described an unusual transverse spin angular momentum of linearly-polarized evanescent (e.g., surface plasmon-polariton) electromagnetic waves [53]. This was followed by paper [69] which claims that such spin is *present* only in transverse-*magnetic* modes and *absent* in transverse-*electric* waves, and even that "the rotation of the magnetic field cannot generate spin". Such misleading conclusions appeared because of the use of the dual-asymmetric definition of the spin density $\tilde{\mathbf{S}}$, $\bar{\tilde{\mathbf{S}}} = \frac{1}{2\omega}\mathrm{Im}\left(\mathbf{E}^* \times \mathbf{E}\right)$, appearing in standard electromagnetism. Obviously, the presence of the angular momentum of the free field should not be attributed solely to the electric rather than magnetic field. The use of the proper dual-symmetric spin density $\mathbf{S}$ and Eq. (3.45) removes this problem, so that both transverse-electric and transverse-magnetic evanescent modes carry the same spin angular momentum [53].

## 5. Concluding remarks

To summarize, we have constructed a classical Lagrangian electromagnetism possessing *dual symmetry* with respect to the electric and magnetic fields. This symmetry is a fundamental property of Maxwell equations which corresponds to the *helicity conservation* law, where the *helicity flux density* coincides with the *spin angular momentum density*. Therefore, we conclude that the dual symmetry is also closely related to the separation of the *spin* and *orbital* degrees of freedom in the electromagnetic field. It is important to note that such separation can only be made using *canonical* Noether currents corresponding to the Poincaré symmetries (i.e., canonical energy-momentum and angular-momentum tensors).

The standard Lagrangian formulation of electromagnetism lacks the dual symmetry, and the helicity conservation is derived in a nontrivial way [12]. Components of the canonical energy-momentum and angular-momentum tensors also lack dual symmetry and contain important discrepancies. In particular, spin density *differs* from the helicity flux density, a *false* (non-conserved) energy density appears in the boost momentum, etc. In addition, both local and integral values of the spin and orbital angular momenta are not dual symmetric, which is bizarre for the free electromagnetic field.

In contrast, the dual electromagnetism suggested in this paper is free of all these drawbacks. The helicity conservation naturally appears here as the basic current from the $U(1)$ gauge transformation. The canonical energy-momentum and angular momentum tensors provide a meaningful and dual-symmetric separation of the spin and orbital degrees of freedom of the field. In particular, the spin density *coincides* with the helicity flux density, the *true* energy density appears in the boost momentum, etc. The spin and orbital momentum and angular-momentum densities following from the dual electromagnetism are in agreement with the expressions suggested recently within several other approaches [3,42–44]. Thus, the dual electromagnetic theory inherently contains straightforward and physically meaningful descriptions of the helicity, spin and orbital characteristics of light. A comparative summary of the main conserved quantities in the standard and dual electromagnetic theories is shown in Fig. 1.

In addition to the formal consideration of the characteristics of the free electromagnetic field, we have discussed their measurability and possible observable consequences of the two theories. It should be taken into account that any measurement of the field characteristics involves *light-matter interactions* and can critically depend on the properties of the *measuring device*. Therefore, the dual symmetry can be *broken* by the measuring device, which is typically sensitive to the *electric* rather than magnetic parts of the optical fields. Understanding of the inherent dual symmetry of the free field and asymmetry of matter offers a powerful tool for analysis of light-matter interactions and suggests clarification and deeper interpretation of a number of experimental and theoretical results [29,30,36,41,53−56,58,59,66−69]. At the same time, the dual asymmetry of measuring devices does *not* mean that one should ascribe dual-asymmetric non-conserved characteristics (e.g., the false energy density $\tilde{W} = E^2$) to the electromagnetic field. Therefore, all fundamental characteristics of the free field must be dual-symmetric, as appears only within the dual electromagnetism.

Importantly, the spin and orbital angular momenta and local energy fluxes are regarded as separably observable quantities in optics. Probe particles move and spin, experiencing the local action of orbital and spin degrees of freedom of the field. If such particles have equivalent electric and magnetic properties, their evolution corresponds to the spin density and orbital energy flux which appear in the dual electromagnetism. Furthermore, the spatial distribution of the field energy density (including both electric and magnetic parts) is directly related to the orbital angular momentum of the field [42,60]. This distribution is consistent with the dual-symmetric orbital angular momentum obtained in our theory rather than with that following from the standard electromagnetic theory.

Thus, it seems that there are grounds to discriminate between the two formulations of electromagnetism in favour of the dual version. This brings up a provocative question about classical field theory: *Can we discriminate between different field Lagrangians leading to the same equations of motion?* The usually assumed answer is "*no*". However, from our consideration it follows that if the spin and orbital angular momenta and local currents are measurable, then the answer is "*yes*". (Different Lagrangians yield different Noether currents and different spin and orbital angular momenta.) A similar question was considered in a recent paper [61], discussing nonrelativistic quantum mechanics and ways to discriminate between different possible definitions of the local probability current. It is argued there that *quantum weak measurements* of the field momentum allows measurements of the local current and enables to single out one particular definition of the current. And, indeed, this was experimentally implemented for an electromagnetic field in [65]. Thus, it seems that modern concepts of quantum measurements and classical field theories enter in contradiction with each other. The probable resolution of this contradiction lies in the separation of the "measured" and "measuring" systems. Light and matter are considered as a single macro-system in field theory, while "matter measures light" in the quantum weak-measurement approach. Apparently, both points of view make physical sense in their corresponding areas of validity.

Finally, we have considered only *classical* electromagnetism. Field quantization and possible manifestations of the dual electromagnetism in *quantum electrodynamics* rise intriguing and nontrivial questions. Interaction with matter must be included in such theory, even if it only appears via virtual particles. Should the matter include magnetic monopoles or there should be a dual-symmetry breaking mechanism? Will this affect observable quantities in quantum electrodynamics, such as atomic-level shifts, particle-scattering cross-section etc? We hope that this paper motivates the analysis of quantum aspects of dual electromagnetism.

We are grateful to Stanley Deser, Yakov Shnir for fruitful and critical correspondence, to Ioannis Bakas, Michael Berry, Yuri Bliokh, Mark Dennis, and Abraham Kofman for helpful discussions, and to Igor Ivanov for bringing Ref. [20] to our attention. This work was partially supported by the European Commission (Marie Curie Action), ARO, JSPS-RFBR contract No. 12-02-92100, Grant-in-Aid for Scientific Research (S), MEXT Kakenhi on Quantum Cybernetics, and the JSPS via its FIRST program.

*Note added.* After posting the first version of this paper in arXiv, another paper by Cameron and Barnett discussing dual-symmetric electromagnetism based on the Lagrangian (3.2) was submitted to the New J. Phys. [70]. This also brought some previous relevant works [71−73] to our attention. Papers [69−72] examine a number of additional conservation laws: Lipkin's zilches and those corresponding to special conformal symmetries. However, they do not treat *canonical* Noether currents corresponding to the Poincaré symmetries, which provide a separation of the spin and orbital degrees of freedom. Therefore, while Cameron and Barnett consider the dual-symmetric formalism as "an alternative rather than a replacement" to the standard approach, we argue that the choice of Lagrangian makes difference and has important physical consequences. Note also that our complex Riemann-Silberstein-like formalism sheds light on the appearance of the "trivial partners" in the conservation laws discussed in [70]. Using the complex potential $\mathrm{X}^\alpha = A^\alpha + i\,C^\alpha$, one can see that the partner conformal transformations (6.18) in [70] represent regular conformal transformations (6.11) but with *imaginary* parameters. In particular, this explains the partner relations between the spin rotation and boost symmetries – a Lorentz boost is a rotation of the Riemann-Silberstein vector by an imaginary angle [38]. Correspondingly, the partner Noether currents are given by the real and imaginary parts of the same complex tensors in our formalism.